\documentclass[useAMS,usenatbib]{mn2e}
\usepackage{aas_macros,graphicx,times,multirow}
\title[X-ray observations of \grb]
  {\xmm\ and \swift\ observations prove \grb\ to be a distant, standard, long GRB}
\author[A. De Luca et al.]
{A.~De~Luca,$^{1,2}$\thanks{E-mail: deluca@iasf-milano.inaf.it}
  P.~Esposito,$^{2,3}$ G.~L.~Israel,$^{4}$ D.~G\"otz,$^{5}$ G.~Novara,$^{2}$   A.~Tiengo$^{2}$\newauthor and S.~Mereghetti$^{2}$\\
$^1$IUSS - Istituto Universitario di Studi Superiori, viale Lungo Ticino Sforza 56, 27100 Pavia, Italy\\
$^2$INAF/Istituto di Astrofisica Spaziale e Fisica Cosmica - Milano, via E.~Bassini 15, 20133 Milano, Italy\\
$^3$Istituto Nazionale di Fisica Nucleare, sezione di Pavia, via A.~Bassi 6, 27100 Pavia, Italy\\
$^4$INAF/Osservatorio Astronomico di Roma, via Frascati 33, 00040 Monteporzio Catone, Italy\\
$^{5}$CEA Saclay, DSM/Irfu/Service d'Astrophysique, Orme des Merisiers, B\^at. 709, 91191 Gif-sur-Yvette, France
}
\date{Accepted 2009 November 09. Received 2009 November 03; in original form 2009 October 15}
\pagerange{\pageref{firstpage}--\pageref{lastpage}} \pubyear{2009}
\def\LaTeX{L\kern-.36em\raise.3ex\hbox{a}\kern-.15em
    T\kern-.1667em\lower.7ex\hbox{E}\kern-.125emX}
\def\xmm {\emph{XMM-Newton}}

\def\swift {\emph{Swift}}

\def\igr {\emph{INTEGRAL}}
\def\grb {GRB\,090709A}

\begin{document}
\label{firstpage}
\maketitle
\begin{abstract}
\grb\ is a long gamma-ray burst (GRB) discovered by \swift, featuring a 
bright X-ray afterglow as well as a faint infrared transient with very 
red and peculiar colors. The burst attracted a large interest because of 
a possible quasi-periodicity at $P=8.1$ s in the prompt emission, suggesting 
that it could have a different origin with respect to standard, long GRBs. 
In order to understand the nature of this burst, we obtained 
a target of opportunity observation with \xmm. X-ray spectroscopy,
based on \xmm\ and \swift\ data, allowed us to model the significant
excess in photoelectric absorption with respect to the Galactic value
as due to a large column density ($\sim$$6.5\times10^{22}$ cm$^{-2}$) in the GRB
host, located at $z\sim4.2$. Such a picture is also consistent with
the infrared transient's properties. Re-analysis of the prompt emission, 
based on \emph{International Gamma-Ray Astrophysics Laboratory} and on \swift\ data, excludes any significant modulation at 
$P=8.1$ s. Thus, we conclude that \grb\ is a distant, standard, long GRB.
\end{abstract}

\section{Introduction}
\begin{figure}
\centering
\resizebox{\hsize}{!}{\includegraphics[angle=-90]{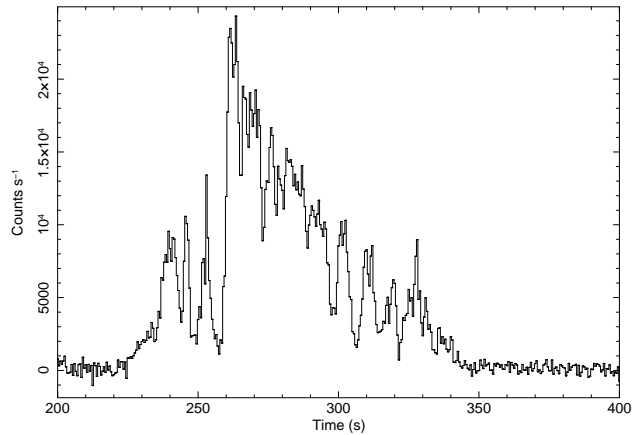}}
\caption{\label{lcbat} Light curve of \grb\ obtained with the \swift/BAT in the 15--350 keV. The time binning is 0.5 s. 
}
\end{figure}

The bright, long gamma-ray burst (GRB) \grb\ was discovered by the Burst Alert Telescope (BAT) 
onboard \swift\ on 2009 July 9 at $T_0=07$:38:34.59 UT \citep{morris09}. The
prompt emission had a complex structure
(see Fig.~\ref{lcbat}), with a multi-peaked light curve
lasting $t_{90}\sim89$ s. The peak flux was $7.8\pm0.3$ ph cm$^{-2}$ s$^{-1}$
at $T_0+21$ s and the fluence was $(2.57\pm0.03)\times10^{-5}$ erg cm$^{-2}$ 
in the 15--350 keV energy range \citep{sakamoto09}.
A bright X-ray afterglow was observed by the X-ray Telescope (XRT) onboard
\swift\, starting as soon as 77 s after the trigger \citep{morris09}. 
Analysis of XRT data up to $T_0+0.3$ days yielded evidence for a break in 
the decay at $T_0+0.1$ days. The XRT spectrum was described by a power law  
with an absorbing column exceeding by a factor $\sim$3 the Galactic value 
in the direction of \grb\ \citep{rowlinson09}.

On the optical/infrared side, several follow-up observations of the 
field of \grb\ were performed. A possible, faint transient with very red 
colors was detected in early observations in the near infrared, 
performed within a few minutes from $T_0$ by automated instruments such as 
the PAIRITEL Telescope \citep*{morgan09}, the Palomar Observatory's 60-inch telescope 
\citep{cenko09} as well as the Faulkes North Telescope \citep{guidorzi09}. 
The peculiar colors of such a transient prompted to suggest a very high 
redshift ($z\sim10$) for such GRB.
The same infrared source was possibly detected in a Subaru image collected 
$\sim$2 hours after the trigger \citep{aoki09}. 
A very deep observation performed with the
10.4 m Gran Telescopio CANARIAS telescope $\sim$41 hours after the GRB
detected no source at the same position, down to
a $3\sigma$ upper limit $I>25.5$ \citep{castrotirado09}.
\\
\indent \grb\ attracted much interest because of a very peculiar
timing phenomenology. \citet{markwardt09}  reported
evidence for quasi-periodical pulsations at $\sim$8 s in the 
prompt emission, based on analysis of \swift/BAT data in the 
15--350 keV energy range. Such a quasi-periodicity
was then confirmed by the analysis of Konus-Wind and Konus-RF data
\citep{golenetskii09}, as well as by analysis of \igr/ACS
data \citep{gotz09}. Taken at face value, such a result is tantalizing.
This would be the first detection of periodic
activity in the prompt gamma-ray emission of a GRB, which could yield
rare information on the behavior of the central engine.\\
\indent Indeed, the peculiar quasi-periodicity immediately prompted
several authors (e.g. \citealt{markwardt09} and \citealt{guidorzi09}) to suggest 
that \grb\ could be different in origin with respect to standard long GRBs, 
and possibly related to activity of a magnetar, either within the Galaxy, 
or extragalactic. However, a search for pulsations in early afterglow data 
collected by \swift/XRT yielded null results \citep{mirabal09}.\\
\indent To shed light on the nature of the peculiar \grb\, we asked 
for an \xmm\ Target of Opportunity (ToO) observation, aimed at
characterizing the afterglow emission at a later stage (about 2 days 
after the event), when the target is too faint to be studied
with \swift/XRT. 
Here we report on the outcome of the \xmm\ ToO, which, coupled
to a re-analysis of \swift\ and \igr\ data, allowed us to obtain
a comprehensive description of the high-energy phenomenology of \grb.

\section{Is the X-ray afterglow ``peculiar''?}
\subsection{The \xmm\ view of \grb\ }
The ToO observation started on 2009 July 11 at 07:56:50 UTC (48.3 hr
after the burst) and lasted 24.5 ks. All the European Photon Imaging Camera (EPIC) detectors were 
operated in Full Frame mode (imaging across the whole field of view,
with a time resolution of 73 ms and 2.6 s in the pn and in the 
two MOS cameras, respectively), using the thin optical filter.
We processed Observation Data Files using the most recent
release of the \xmm\ Science Analysis Software 
(\textsc{sas}v9.0).\\
\indent No significant particle background episodes affected
the observation. The afterglow is clearly detected 
in all of the EPIC cameras. The position, $\rm RA=19^h19^m42\fs6$,
$\rm Dec.= +60\degr43\arcmin35\farcs0$ (J2000), with a $1\sigma$ error 
of $1\farcs5$, is consistent with the XRT position \citep{osborne09}.\\
\indent Source photons were selected from a circular region 
(30 arcsec radius) centered  on the target. Background events were 
extracted from a source-free region in the same chip as the target.
The source average background-subtracted count rate in the 0.3--8 keV 
energy range is $0.155\pm0.003$ counts s$^{-1}$, $0.051\pm0.002$ 
counts s$^{-1}$ and $0.048\pm0.002$ counts s$^{-1}$
in the pn, MOS1 and MOS2 cameras, respectively. Background
contributes $\sim$5\% more counts in the extraction region.

\subsubsection{Search for pulsations}
\begin{figure}
\resizebox{\hsize}{!}{\includegraphics[angle=0]{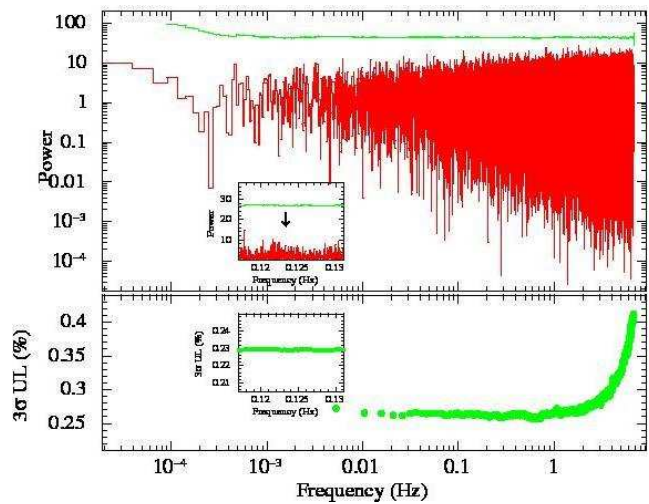}}
\caption{\label{pntim} Results of the timing analysis on EPIC/pn data (0.5--10
  keV), collected $\sim$48 hours after the trigger. Upper panel: the power spectrum is shown 
together with the threshold for the detection of sinusoidal signals at 
the 3$\sigma$ confidence level. Lower panel: upper limits on the pulsed
fraction. See text for details.}
\end{figure}

As a first step, after correcting photon arrival times 
to the solar system's barycenter using the task \textsc{barycen},
we searched for pulsations in the GRB afterglow using two methods. 
First, we used the $Z_n^2$ test \citep{buccheri83}, with 
the number of harmonics $n$ being varied from 1 to 2. 
Second, a Fourier analysis of the light-curves was performed using
the method described in \citet{israel96}. The analysed
period range spans from 150\,ms up to 10$^4$\,s ($\sim$262\,000 total period 
trials). 
No significant (periodic or quasi-periodic) signal was found with either method, searching the
whole period range or restricting the search around the 8 s hypothetical
period. The upper limits on the pulsed fraction, computed
according to \citet{vaughan94}, are shown in Figure 2 as a function
of the frequency (the curves refer to the 3$\sigma$ upper limits on the
sinusoid semi-amplitude pulsed fraction in the 0.5--10 keV energy range). 
Upper limits at the 8 s hypothetical period range between 26\% when 
considering all the Fourier frequencies in the whole spectrum and 23\% 
for the narrower search (see insets in Figure~\ref{pntim}, lower panel).

\subsubsection{Spectral analysis}
Source spectra were rebinned in order to oversample the instrumental 
energetic resolution by a maximum factor 3, or to have at 
least 25 counts per bin. Ad-hoc response matrices and effective area files 
were generated using the \textsc{sas} tasks \textsc{rmfgen} and 
\textsc{arfgen}, respectively.\\
\indent Spectral analysis was performed using the \textsc{xspec} software 
(v12.4.0). Errors on spectral parameters 
are given at the 90\% confidence level for a single parameter of interest.
Simultaneous modelling of pn, MOS1 and MOS2 spectra with an
absorbed power law model results in a rather poor fit ($\chi^2_{\nu}=1.52$,
152 d.o.f.). The best fit photon index is $\Gamma=2.15\pm0.05$, while the
absorbing column ($N_{\rm H}=(2.5\pm0.2)\times10^{21}$ cm$^{-2}$) is significantly 
larger than the Galactic value in the direction of \grb\ 
\citep[$\sim$$6.7\times10^{20}$ cm$^{-2}$, according to][]{dickey90}.
Such result are in broad agreement with the XRT ones reported by 
\citet{rowlinson09}.\\
\indent We then added to the Galactic absorption component (with 
$N_{\rm H}$ fixed to $6.7\times10^{20}$ cm$^{-2}$) a redshifted 
absorption component $N_{\mathrm{H},z}$ 
(abundances were set to Solar values). 
This model yields a 
much better description of the data ($\chi^2_{\nu}=1.11$,
151 d.o.f.). The resulting photon index is $\Gamma=2.00\pm0.05$,
The fit gives a very large 
intrinsic column density $N_{\mathrm{H},z}=(9.7\pm2.0)\times10^{22}$ cm$^{-2}$,
while the redshift is constrained to $5.1\pm0.4$. 
The time-averaged, observed flux is $5.6\times10^{-13}$ erg cm$^{-2}$ s$^{-1}$
in the 0.5--10 keV energy range.

\subsection{Swift/XRT data}
In order to put the \xmm\ observation in context, we retrieved and
analysed the XRT observations of the afterglow. 
The XRT CCD detector (0.2--10 keV) started observing the field of \grb\ only
77 s after the BAT trigger. Table \ref{log} reports the log of this and the subsequent XRT follow-up observations, performed in both photon counting (pc) and windowed timing (wt) modes,\footnote{In pc mode the entire CCD is read every 2.507 s, while in wt mode only the central 200 columns are read and only one-dimensional imaging is preserved, achieving a time resolution of 1.766 ms (see \citealt{hill04short} for more details).} that were used for this work (some wt observations last only a few seconds and we did not include them in our analysis).\\
\indent The data were processed with standard procedures (\textsc{xrtpipeline} version 0.12.3), filtering, and screening criteria by using \textsc{ftools} in the \textsc{heasoft} package (ver. 6.6). For the timing and spectral analyses, we extracted the pc source events from a circle with a radius of 20 pixels (one pixel corresponds to about $2\farcs36$) and the wt data from a $40\times40$ pixels box along the image strip. To estimate the background, we extracted pc and wt events from source-free regions far from the position of \grb.\\
\begin{figure}
\centering
\resizebox{\hsize}{!}{\includegraphics[angle=-90]{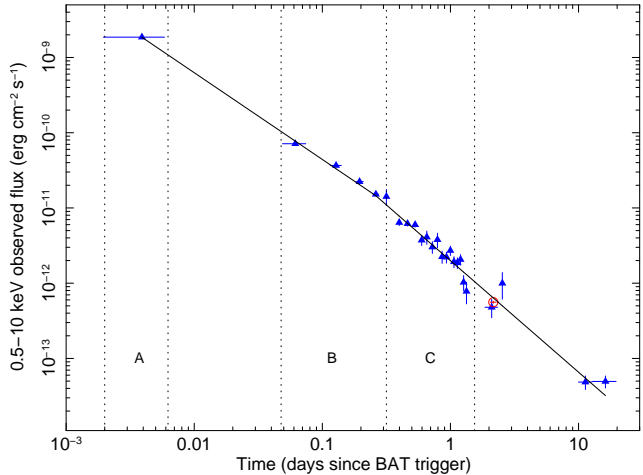}}
\caption{\label{lc} Soft X-ray light-curve of \grb spanning the time range
from $T_0+77$ s to $T_0+20$ days. Blue triangles: \swift/XRT data; Red
  circle: \xmm\ data. The broken power law best describing the flux decay
is superimposed. Time intervals for XRT time-resolved spectroscopy are marked
as A, B, and C.}
\end{figure}
\begin{table*}
\centering
\begin{minipage}{12.5cm}
\caption{Journal of the \swift/XRT observations. The \swift/BAT trigger time is 2009-07-09 07:38:35 UT.}
\label{log}
\begin{tabular}{@{}ccccc}
\hline
Sequence/Mode & \multicolumn{2}{c}{Start/End time (UT)} & Exposure$^{a}$ & Net average count rate$^{b}$\\
  & \multicolumn{2}{c}{yyyy-mm-dd hh:mm:ss} & (ks) & (counts s$^{-1}$)\\
\hline
00356890000/wt & 2009-07-09 07:39:52 & 2009-07-09 07:47:33 & 0.4  & $35.4\pm0.3$\\
00356890000/pc & 2009-07-09 08:47:09 & 2009-07-09 15:14:03 & 8.9  & $0.744\pm0.009$\\
00356890001/pc & 2009-07-09 16:56:40 & 2009-07-09 02:29:10 & 7.9  & $0.103\pm0.004$\\
00356890002/pc & 2009-07-10 02:34:55 & 2009-07-10 20:40:56 & 10.2 & $0.036\pm0.002$\\
00356890259/wt & 2009-07-11 04:14:24 & 2009-07-11 15:39:27 & 12.0 & $0.008\pm0.003$\\
00356890003-4-5/pc & 2009-07-19 06:56:06 & 2009-07-21 23:38:57 & 30.4 & $0.0011\pm0.0002$\\
00356890006-7-9-10-11/pc & 2009-07-22 02:39:01 & 2009-07-28 22:45:58 & 53.4 & $0.0012\pm0.0002$\\
\hline
\end{tabular}
\begin{list}{}{}
\item[$^{a}$] The exposure time is usually spread over several snapshots (single continuous pointings at the target) during each observation.
\item[$^{b}$] In the 0.5--10 keV energy band.
\end{list}
\end{minipage}
\end{table*}
\begin{table*}
\centering
\begin{minipage}{12.5cm}
\caption{Results of time-resolved spectroscopy with \swift/XRT and XMM-Newton/EPIC.}
\label{spec}
\begin{tabular}{@{}ccccc}
\hline
 & \swift/XRT A & \swift/XRT B & \swift/XRT C & \xmm/EPIC \\
\hline
Exp.Time (ks) & 0.4 & 8.9 & 17.9 & 18.5$^a$/22.6$^b$ \\
Mean epoch (MJD) & 55021.322 & 55021.500 & 55022.248 & 55023.480 \\
$N_{\rm H}$ (cm$^{-2}$, fixed) & $6.7\times10^{20}$ & $6.7\times10^{20}$ &
$6.7\times10^{20}$ & $6.7\times10^{20}$ \\
$N_{\mathrm{H},z}$ (cm$^{-2}$) & $(6.2\pm1.1)\times10^{22}$ & $(8.7\pm1.7)\times10^{22}$
& $(7.1^{+6.8}_{-3.5})\times10^{22}$ & $(9.7\pm2.2)\times10^{22}$ \\
$z$ & $4.1\pm0.3$ & $4.3\pm0.3$ & $3.9\pm1.0$ & $5.1\pm0.5$ \\
$\Gamma$ & $1.69\pm0.03$ &  $1.84\pm0.05$ & $2.05\pm0.14$ & $2.00\pm0.06$ \\
$\chi^2_{\nu}$ & 1.13 & 0.98 & 1.12 & 1.12 \\
d.o.f. & 291 & 172 & 39 & 151 \\
F$^c$ (erg cm$^{-2}$ s$^{-1}$) & $1.6\times10^{-9}$ & $3.7\times10^{-11}$ &  $1.6\times10^{-12}$ & $5.6\times10^{-13}$ \\
\hline
\end{tabular}
\begin{list}{}{}
\item[$^{a}$/$^{b}$] pn/MOS.
\item[$^{c}$] Observed flux in the 0.5--10 keV energy band.
\end{list}
\end{minipage}
\end{table*}
\indent The overall light curve of the afterglow of \grb\ is shown in 
Fig.~\ref{lc}. Count rates (for both \swift/XRT and \xmm/EPIC) have been 
converted  into fluxes using the best-fitting spectral models described 
below and in the previous section. The flux decays as a broken power law. 
The break occurs at $0.26\pm0.05$ days after the GRB, when the index of the 
decay changes from $-1.15\pm0.01$ to $-1.48\pm0.05$. 

\subsubsection{Search for pulsations}
We calculated the power 
spectrum for the first \swift\ dataset during which the time resolution 
was higher (WT mode) and with large enough statistics. The analysed
period range spans from 4 ms up to 10$^3$ s ($\sim$262\,000 total period 
trial) approximatively. Also in this case no significant signal was found 
searching the whole period range or restricting the search around the 8 s 
signal.  Due to the presence of low-frequency noise (introduced by the 
source rapid decay during the \swift\ observation), meaningful upper limits 
in the range 60\%--50\% have been inferred only for the narrow search. The 
subtraction of high-order de-trending polynomials (taking account for the source 
rapid decay) does not change significantely  the above results. We note the 
presence of a low-significance ($\sim$2$\sigma$) QPO-like feature in the 
power spectrum around 11 s (Fig.~4). 

\begin{figure}
\resizebox{\hsize}{!}{\includegraphics[angle=0]{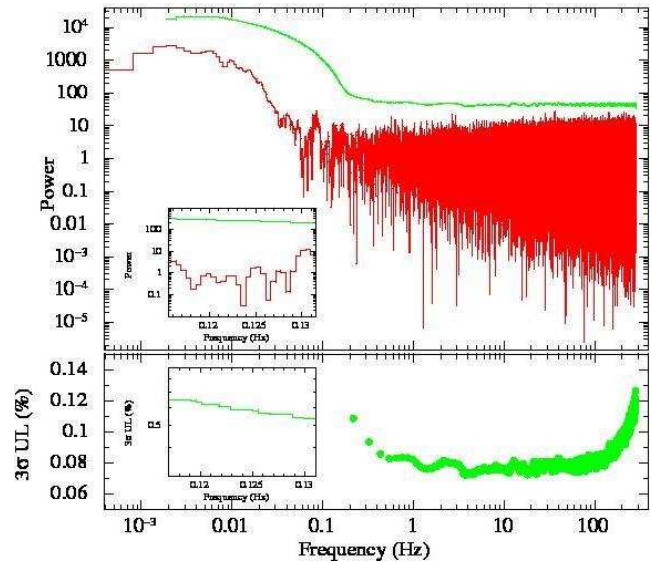}}
\caption{\label{xrttim} Same as Fig.~\ref{pntim}, using the data collected with
\swift/XRT (0.5--10 keV) in WT mode in the early afterglow phase (from 
$T_0+77$ s to $T_0+539$ s).}
\end{figure}

\subsubsection{Spectral analysis}
\label{spectra}
\begin{figure}
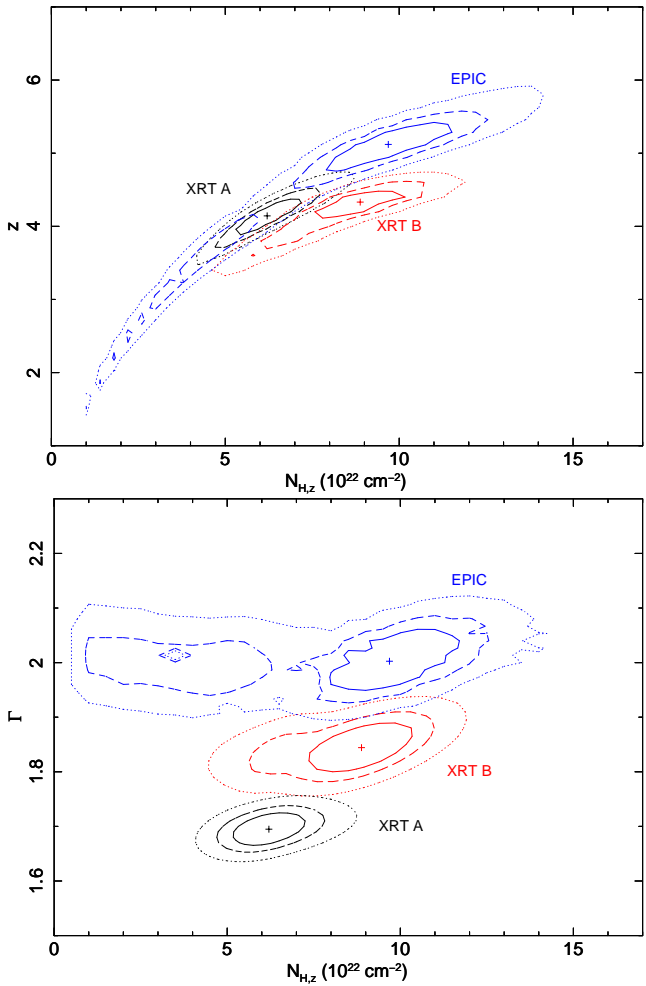

\centering
\resizebox{\hsize}{!}{\includegraphics[angle=-90]{nhz-z-tot.eps}}
\resizebox{\hsize}{!}{\includegraphics[angle=-90]{nhz-gamma-tot.eps}}
\caption{\label{gamma-nhz}Upper panel: confidence contours (solid: 68\%, dashed: 90\%,
dotted: 99\%) for the redshift $z$ vs. the intrinsic column density $N_{\mathrm{H},z}$. Lower panel: confidence contours (solid: 68\%, dashed: 90\%,
dotted: 99\%) for the photon index $\Gamma$ vs. the intrinsic column density $N_{\mathrm{H},z}$.}
\end{figure}

We extracted time-resolved spectra of the afterglow. 
In order to have a good photon statistics, we combined data
collected within the three time intervals marked as
A, B, and C in Fig.~\ref{lc}.\\
\indent A simple, absorbed power-law model does not reproduce well the 
spectra and results in a $N_{\rm H}$ column significantly larger than the 
Galactic value. As in the \xmm\ case, we added a redshifted absorption 
component, which yielded a much better fit to the data. Results are reported 
in Table~\ref{spec}.\\
\indent In order to assess the possible spectral evolution as a function of
time, we generated confidence ellipses for $z$ vs. $N_{\mathrm{H},z}$ and for
$\Gamma$ vs.  $N_{\mathrm{H},z}$ for the three datasets having the largest 
photon statistics  (XRT A, XRT B and EPIC). As shown in Fig.~\ref{gamma-nhz}, 
the strong correlation between the $N_{\mathrm{H},z}$ 
and $z$ parameters prevents from concluding that we are observing different 
values in different epochs. Conversely, a significant softening of the 
power law component is apparent. The photon index varies from $\sim$1.7 in 
the early afterglow phase to $\sim$2 at the time of the \xmm\ observation.\\
\indent Indeed, fitting the above model simultaneously to all XRT and EPIC 
spectra,  leaving the power law as the only spectral component free to vary 
as a function of the epoch, yields a very good result ($\chi^2_{\nu}=1.08$, 
671 d.o.f.). Such an exercise yields a best fit redshift $z=4.2\pm0.2$ and 
intrinsic absorption $N_{\mathrm{H},z}=(6.5\pm1.5)\times10^{22}$ cm$^{-2}$, 
while the epoch-dependent values of $\Gamma$ and of the power-law 
normalization  are very similar to the ones reported in Table~\ref{spec}.
Taking into account the correlation between $z$ and $N_{\mathrm{H},z}$,
  as well as the dependence of such parameters on the value of the Milky Way
  column density\footnote{\citet{stratta04} showed that a reasonable guess
of the uncertainty affecting Galactic
column density estimates is $\sim30\%$ at $\sim90\%$ confidence level.}, 
the 90\% confidence level interval turn out to
  be $3.7<z<4.5$ and
  $4.8\times10^{22}\,\mathrm{cm}^{-2}<N_{\mathrm{H},z}<7.7\times10^{22}\,\mathrm{cm}^{-2}$. We
  also note that such results are computed assuming Solar abundances in the
  redshifted absorber model.

\section{Any periodicity in the prompt emission?}
Facing with a rather standard X-ray afterglow, we
have studied the prompt emission in order to assess the 
significance of the claimed quasi-periodical variations.

\subsection{\emph{\swift/BAT data}}
\begin{figure}
\centering
\resizebox{\hsize}{!}{\includegraphics[angle=-90]{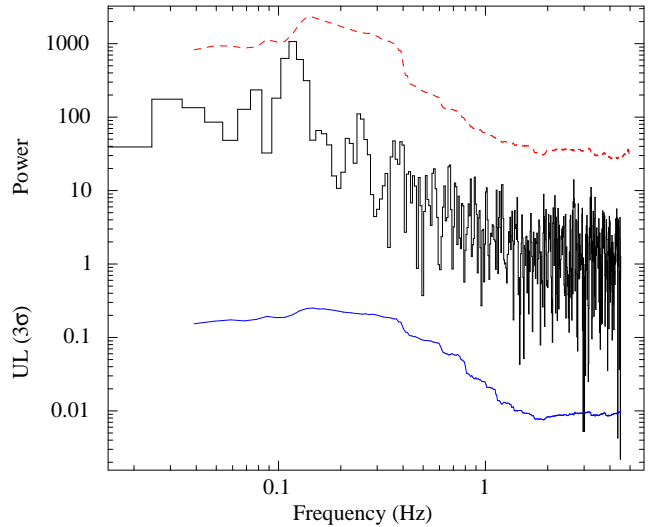}}
\caption{\label{battim} Similar to Fig.~\ref{pntim} for the prompt emission,
using \swift/BAT data (15--150 keV) in the time range from $T_0$ s 
to $T_0+100$ s. The uppermost  
line (stepped) marks the threshold for the detection of sinusoidal signals at 
the 3$\sigma$ c.l., while the lower one shows the corresponding upper limits on 
the pulsation amplitude.}
\end{figure}

The coded-mask Burst Alert Telescope (BAT) onboard \swift\ collected 
event data (15--350 keV, with 0.1 ms time resolution) from \grb\ in the
time range from $T_0-170$ s to $T_0+612$ s. We retrieved the event list 
from the \swift\ archive. After correcting photons' time of arrival to the
solar system barycenter, we generated a background-subtracted light curve
in the 15--150 keV energy range using the mask-weighting technique 
\citep{senziani07}. Then, we used a procedure similar to that already 
discussed for \xmm\ and \swift/XRT data. We searched for possible signals in 
the power spectrum. In this case we considered the whole set of Fourier 
frequencies. We searched for significant signals in the range from 20 ms to 
50 s, approximately. We also included high-order de-trending polynomials to 
``rectify'' (trend subtraction mode was selected) the light curve and minimize 
the low-frequency noise. 
Fig.~\ref{battim} shows our best-case power spectrum, obtained after 
de-trending the light curve by using a third order polynomial. No significant 
signal was found, while a 3$\sigma$ upper limits in the 15\%--25\% range was 
obtained in the period interval between 5 s and 15 s. We note that a peak  at about 
0.125\,Hz, consistent to the reported period of 8\,s, is present in the spectrum. The  
corresponding power estimate is very close to the 3$\sigma$ detection threshold.

We also studied the time-averaged spectrum of the prompt emission,
generating a background-subtracted spectrum with the mask-weighting technique
\citep{senziani07}, considering the time interval from $T_0$ to $T_0+150$ s.
The best fit model  
is a power law with photon index $\Gamma=1.28\pm0.02$. The 
fluence in the 15--150 keV energy range is $2.38\times10^{-5}$  erg cm$^{-2}$. Our results are fully 
consistent to the ones by \citet{sakamoto09}.

\subsection{\emph{INTEGRAL SPI/ACS data}}
\begin{figure}
\centering
\resizebox{\hsize}{!}{\includegraphics[angle=-90]{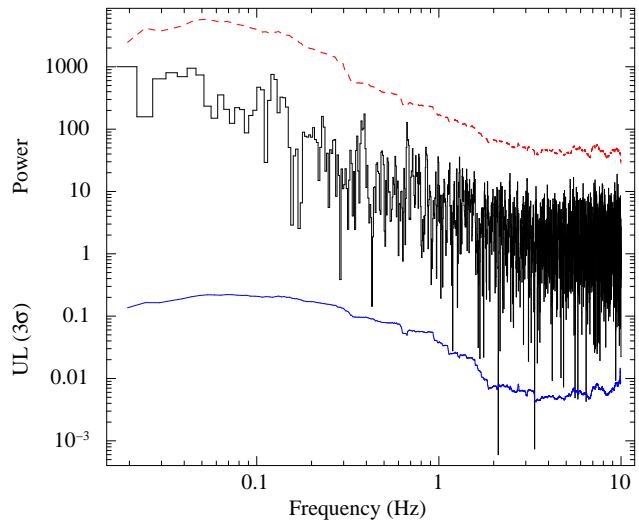}}
\caption{\label{fig:timing-acs} Similar to Fig.~\ref{battim} but for the prompt emission as 
seen by \igr/SPI-ACS. The uppermost line (stepped) marks the threshold for the detection of 
sinusoidal signals at the 3$\sigma$ c.l., while the lower one shows the corresponding 
upper limits on  the pulsation amplitude. }
\end{figure}

The Anti-Coincidence Shield (ACS) of the Spectrometer on \igr\ (SPI) is routinely 
used as a nearly omni-directional detector for gamma-ray bursts \citep{vonkienlin03short}, 
besides serving its main function as a veto for the SPI spectrometer. The ACS provides 
light curves binned at 50 ms, but without energy and directional information. The low energy 
threshold is about 80 keV.\\
\indent \grb\ was located at an angle $\theta=70\degr$ from the SPI pointing direction, 
resulting in an optimal response for the ACS, which is most sensitive for directions 
orthogonal to the satellite pointing axis.\\
\indent In order to look for a possible periodic signal in the ACS data we
perfomed the same analysis as in the \swift/BAT case (see above). The results are 
shown in Fig. \ref{fig:timing-acs}: also in this case the search was negative. 
The 3$\sigma$ upper limits to any pulsed signal around 0.125 Hz is 20\%, and 6\% and 8\% 
for the blind search and the narrow search in the 5--15\,s period interval, respectively.

\section{Discussion and conclusions}\label{disc}
\subsection{The afterglow of a distant, standard, long GRB}
Prompted by early reports of the detection of a strong quasi-periodical 
signal in the prompt emission of \grb, we have performed a ToO observation 
with \xmm.\\ 
\indent The afterglow of \grb\, as seen in soft X-rays with the EPIC 
instrument, is fully similar to other cases of typical, well-behaving,
long GRBs. No pulsations are seen in the X-ray emission $\sim$48 hr
after the trigger, with a $3\sigma$ upper limit of 15--25\%
on the pulsed fraction, assuming a sinusoidal pulse shape.
The energy spectrum has a typical power-law shape, with a remarkable 
excess in photoelectric absorption (by a factor $\sim$4) 
with respect to the Galactic $N_{\rm H}$ in the direction of \grb.\\ 
\indent Such a picture of a ``standard'' X-ray afterglow is completed 
by \swift/XRT observations, which started as soon as 77 s after the burst, 
and extended up to 20 days after the trigger. \grb\ turns out to have a rich 
phenomenology, featuring -  in addition to the intrinsic photoeletric 
absorption - a clear spectral evolution (the photon index $\Gamma$ steepens 
from  1.7 to 2.0 in $\sim$2 days), as well as a break in the flux decay, with 
a $\sim0.3$ variation in the power law decay index occurring $\sim$0.26 
days after the burst.\\
\indent We will focus here on the extra absorption only, the evidence for 
which is fully confirmed - and strenghtened - by \swift\ data. Indeed,
adopting a simple, redshifted neutral absorption model, all datasets yield 
consistent values for both the redshift and the redshifted absorbing column, 
as well as of the spectral steepening as a function of time. \xmm\ and 
\swift/XRT data point  to a rather high redshift ($z\sim4.2$) for \grb\ as 
well as to a huge column density in the host galaxy 
($N_{\rm H}\sim6\times10^{22}$ cm$^{-2}$). 
Although a word of caution is required in considering such results (see Sect.~\ref{spectra}),
to our knowledge, this is the 
largest intrinsic absorption ever observed in an X-ray afterglow spectrum 
\citep[see e.g.][]{deluca05,campana06short,grupe07}.\\
\indent This high absorption is also on the high side among expected column 
densities for sources located inside molecular clouds \citep{reichart02}.
We note that part of the absorption 
could be due to intervening systems along the line of sight rather than
being local to the host galaxy, as seen in the optical band in QSO and GRB
studies 
\citep[the so-called Damped Lyman-$\alpha$ Absorbers;][]{reichart02,wolfe05}. 
Lack of optical spectroscopy in the case of \grb\ prevents from drawing firm 
conclusions.\\
\indent Such a picture suggests an explanation of the peculiar, very red 
colors of the infrared transient likely associated to \grb\ \citep{morgan09}
as due to  reddening in the host galaxy. Indeed, if the GRB spectrum has no 
breaks in the X-ray to optical range 
and the redshift and intrinsic column 
derived from X-ray spectroscopy are correct, 
an extinction
A$_V\sim3$ in the GRB host galaxy (at $z=4.2$) would fit the infrared data. 
This would point to a dust-to-gas ratio $\sim$10\% of that found in the 
Milky Way (assuming Solar abundances), similar to the findings of
other investigations \citep[e.g. ][]{galama01,hjorth03short}. Such an 
interpretation of the infrared data is possibly supported by the
time decay of the infrared transient. The fading between 
the PAIRITEL and the Subaru observations is consistent
with the $t^{-1.15}$ law describing the X-ray afterglow decay before 
the break at $T_0+0.26$ days, which suggests a common origin
for the X-ray and infrared emission.\\
\indent In any case, 
 X-ray spectroscopy yields a robust indication that \grb\ was 
a very distant event.

\subsection{On the prompt temporal variability}

Coming back to the temporal properties of the prompt emission, our reanalysis, based on both \swift/BAT and \igr\ SPI/ACS data, could not confirm the presence of any significant signal at $\sim$8.1 s. In any case, a complex multi-peaked light curve is clearly apparent, with a peak-to-peak separation of order $\sim$8--10 s, and a peak at 0.125\,Hz is present in the power spectrum derived from BAT data, just below the 3$\sigma$ detection threshold. It is premature to comment on the apparent discrepancy between our results and those reported in a circular by \citet{markwardt09}, in view of the lack of detailed information concerning the analysis performed by these authors. Indeed, several issues affect estimation of the significance of a signal when dealing with de-trending algorithms and Fourier transform techniques. Among the most important, we remember the presence of non-Poissonian noise in the power spectrum and the change in the statistical properties of a time series induced by operations such as subtraction and division. The former, if not taken into account properly, may result in an overestimation of the statistical significance of any peak sitting on a non-Poissonian underlying power spectrum continuum (see \citealt{israel96}). The latter affects the statistical properties of the original time-series. In particular, the division of a Possonian variable (the time series) with a (model-dependent) de-trending algorithm might result in a non-Poissonian distribuited time series with effects on the (unknown) statistical properties of the power spectrum noise.\\
\indent The claimed period of about 8 s is precisely in the range of periods of
soft gamma-ray repeaters (SGRs) and anomalous X-ray pulsars (AXPs). These sources are 
thought to be magnetars: isolated neutron stars powered by strong 
(10$^{14}$--10$^{15}$ G) magnetic fields (see \citealt{mereghetti08} for a review). 
Thus, association of \grb\ with the activity of a magnetar was suggested, also 
possibly based on some similarity of the light curve to the one of a magnetar 
giant flare, featuring a sort of ``pulsating tail''.\\
\indent Based on high energy properties, we can rule out such an 
interpretation. First, we could find no unambiguous, coherent pulsations in 
the $\gamma$-ray emission. Then, the spectrum of \grb\ is much harder than 
the typical spectra observed in the pulsating tails of the giant flares from 
SGRs. These are well described by thermal bremsstrahlung emission with
$kT\sim15$--30 keV,  while the spectrum of \grb\  is typical for a GRB.  Giant 
flares are also characterized by a short ($<$0.5 s), very bright and 
spectrally hard initial spike that was clearly absent in \grb\  (although 
this feature was lacking in some ``intermediate flares'' from SGRs, all of 
them had a very sharp initial rise, contrary to the light curve of \grb ). 
The location at high Galactic latitude ($b\sim20\degr$) would also be very 
unusual for a Galactic SGR. Moreover, a Galactic origin may 
be safely ruled out by the observation of an absorbing column
exceeding by a factor $\sim$4 the Galactic value in the direction of \grb.
A giant flare from an extragalactic SGR can also be
excluded, based on the absence of a visible nearby galaxy at the location
of \grb. The observed fluence would imply a distance not 
larger than 600 kpc (assuming an energetic similar to the one of
other observed magnetar giant flares). The enormous
energy requirement implied by a cosmological distance
rules out the SGR giant flare hypothesis.\\
\indent As already discussed in the previous section, we conclude that 
\grb\ is a standard, long GRB, with a 
multi-peak structure in the prompt emission. We will not go into pure
speculations by considering physical processes that could produce a true 
quasi-periodic signal.\\
\indent The variable prompt emission could be the signature of non-stationary 
processes in the GRB inner engine.  As a likely  possibility -- discussed 
by \citet{beskin09} who observed a similar phenomenology in the optical 
emission from the ``naked-eye'' GRB\,080319B (but with a possibly stronger 
evidence for quasi-periodicity) -- the peculiar variability could be related 
to cyclic accretion by the central newborn compact object. Such a phenomenon
could be due to the fragmentation of an accretion disc due to 
some kind of instability. 
For instance, \citet{masada07} explain short-time
variability in the prompt emission by GRBs as due to
magneto-rotational instabilities developing in a massive, hot 
hyperaccreting disc surrounding a central black hole
of a few stellar masses. At a redshift $z\sim4.2$, the variability
time scale in the source frame would be of $\sim1.5$ s. Thus, adopting
the model of \citet{masada07},
the observed properties of
\grb\ could fit into such a scenario by assuming a beaming factor
$\sim$100, for a $\sim$1
$M_{\odot}$ accretion disc with inner radius of $\sim$30
gravitational radii, surrounding 
a $\sim$$4M_{\odot}$ central black hole.


\section*{acknowledgements}
This research is based on observations obtained with \xmm\ and \igr, which are both ESA science missions with instruments and contributions directly funded by ESA Member States and the USA (through NASA), and on observations with the NASA/UK/ASI \swift\ mission. We thank Norbert Schartel and the staff of the \xmm\ Science Operation Center for performing the Target of Opportunity observation. The Italian authors acknowledge the partial support from ASI (ASI/INAF contracts I/088/06/0). DG acknowledges the CNES for financial support.
\bibliographystyle{mn2e}
\bibliography{biblio}

\begin{thebibliography}{}

\bibitem[\protect\citeauthoryear{{Aoki}, {Ishii}, {Kuzuhara}, {Takahashi} \&
  {Kawai}}{{Aoki} et~al.}{2009}]{aoki09}
{Aoki} K.,  {Ishii} M.,  {Kuzuhara} M.,  {Takahashi} Y.,    {Kawai} N.,  2009,
  GCN Circ., 9634

\bibitem[\protect\citeauthoryear{{Beskin}, {Karpov}, {Bondar}, {Guarnieri},
  {Bartolini}, {Greco} \& {Piccioni}}{{Beskin} et~al.}{2009}]{beskin09}
{Beskin} G.,  {Karpov} S.,  {Bondar} S.,  {Guarnieri} A.,  {Bartolini} C.,
  {Greco} G.,    {Piccioni} A.,  2009, submitted to Science, preprint (arXiv:
  astro-ph/0905.4431)

\bibitem[\protect\citeauthoryear{{Buccheri}, {Bennett}, {Bignami}, {Bloemen},
  {Boriakoff}, {Caraveo}, {Hermsen}, {Kanbach}, {Manchester}, {Masnou},
  {Mayer-Hasselwander}, {Ozel}, {Paul}, {Sacco}, {Scarsi} \&
  {Strong}}{{Buccheri} et~al.}{1983}]{buccheri83}
{Buccheri} R. et al.,  1983, \aap, 128, 245

\bibitem[\protect\citeauthoryear{{Campana}, {Romano}, {Covino}, {Lazzati}, {de
  Luca}, {Chincarini}, {Moretti}, {Tagliaferri} \& {et~al.}}{{Campana}
  et~al.}{2006}]{campana06short}
{Campana} S. et~al., 2006, \aap,
  449, 61

\bibitem[\protect\citeauthoryear{{Castro-Tirado}, {de Ugarte Postigo},
  {Gorosabel}, {Guziy}, {Jelinek}, {Kubanek}, {Perez-Ramirez}, {Mirabal},
  {Llorente}, {Castro Ceron}, {Alvarez} \& {Cepa}}{{Castro-Tirado}
  et~al.}{2009}]{castrotirado09}
{Castro-Tirado} A.~J. et al.,  2009, GCN Circ.,
  9655

\bibitem[\protect\citeauthoryear{{Cenko}, {Bloom}, {Morgan} \&
  {Perley}}{{Cenko} et~al.}{2009}]{cenko09}
{Cenko} S.~B.,  {Bloom} J.~S.,  {Morgan} A.~N.,    {Perley} D.~A.,  2009, GCN
  Circ., 9646

\bibitem[\protect\citeauthoryear{{De Luca}, {Melandri}, {Caraveo}, {G{\"o}tz},
  {Mereghetti}, {Tiengo}, {Antonelli}, {Campana}, {Chincarini}, {Covino},
  {D'Avanzo}, {Fernandez-Soto}, {Fugazza}, {Malesani}, {Stella} \&
  {Tagliaferri}}{{de Luca} et~al.}{2005}]{deluca05}
{De Luca} A. et al., 2005, \aap, 440, 85

\bibitem[\protect\citeauthoryear{{Dickey} \& {Lockman}}{{Dickey} \&
  {Lockman}}{1990}]{dickey90}
{Dickey} J.~M.,  {Lockman} F.~J.,  1990, \araa, 28, 215

\bibitem[\protect\citeauthoryear{{Galama} \& {Wijers}}{{Galama} \&
  {Wijers}}{2001}]{galama01}
{Galama} T.~J.,  {Wijers} R.~A.~M.~J.,  2001, \apjl, 549, L209

\bibitem[\protect\citeauthoryear{{Golenetskii}, {Aptekar}, {Mazets},
  {Pal'Shin}, {Frederiks}, {Oleynik}, {Ulanov} \& {Svinkin}}{{Golenetskii}
  et~al.}{2009}]{golenetskii09}
{Golenetskii} S.,  {Aptekar} R.,  {Mazets} E.,  {Pal'Shin} V.,  {Frederiks} D.,
   {Oleynik} P.,  {Ulanov} M.,    {Svinkin} D.,  2009, GCN Circ., 9647

\bibitem[\protect\citeauthoryear{{G\"otz}, {Mereghetti}, {von Kienlin} \&
  {Beck}}{{G\"otz} et~al.}{2009}]{gotz09}
{G\"otz} D.,  {Mereghetti} S.,  {von Kienlin} A.,    {Beck} M.,  2009, GCN Circ.,
  9649

\bibitem[\protect\citeauthoryear{Grupe et al.}{2007}]{grupe07} 
Grupe D., Nousek J.~A., vanden Berk D.~E., Roming P.~W.~A., Burrows D.~N., 
Godet O., Osborne J., Gehrels N., 2007, AJ, 133, 2216 

\bibitem[\protect\citeauthoryear{{Guidorzi}, {Melandri}, {Mundell}, {Bersier},
  {Cano}, {Kobayashi}, {Steele}, {Smith} \& {Gomboc}}{{Guidorzi}
  et~al.}{2009}]{guidorzi09}
{Guidorzi} C. et al.,  2009, GCN
  Circ., 9648

\bibitem[\protect\citeauthoryear{{Hill}, {Burrows}, {Nousek}, {Abbey},
  {Ambrosi}, {Br{\"a}uninger}, {Burkert}, {Campana} \& {et~al.}}{{Hill}
  et~al.}{2004}]{hill04short}
{Hill} J.~E. et~al., 2004, in {Flanagan} K.~A.,  {Siegmund} O.~H.~W.,  eds, X-ray and Gamma-ray
  instrumentation for Astronomy XIII. Vol.~5165 of SPIE Conf. Ser.,
  Bellingham WA, p 217

\bibitem[\protect\citeauthoryear{{Hjorth}, {M{\o}ller}, {Gorosabel}, {Fynbo},
  {Toft}, {Jaunsen}, {Kaas}, {Pursimo} \& {et~al.}}{{Hjorth}
  et~al.}{2003}]{hjorth03short}
{Hjorth} J. et~al., 2003, \apj, 597,
  699

\bibitem[\protect\citeauthoryear{{Israel} \& {Stella}}{{Israel} \&
  {Stella}}{1996}]{israel96}
{Israel} G.~L.,  {Stella} L.,  1996, \apj, 468, 369

\bibitem[\protect\citeauthoryear{{Markwardt}, {Gavriil}, {Palmer},
  {Baumgartner} \& {Barthelmy}}{{Markwardt} et~al.}{2009}]{markwardt09}
{Markwardt} C.~B.,  {Gavriil} F.~P.,  {Palmer} D.~M.,  {Baumgartner} W.~H.,
  {Barthelmy} S.~D.,  2009, GCN Circ., 9645

\bibitem[\protect\citeauthoryear{{Masada}, {Kawanaka}, {Sano} \&
  {Shibata}}{{Masada} et~al.}{2007}]{masada07}
{Masada} Y.,  {Kawanaka} N.,  {Sano} T.,    {Shibata} K.,  2007, \apj, 663, 437

\bibitem[\protect\citeauthoryear{{Mereghetti}}{{Mereghetti}}{2008}]{mereghetti%
08}
{Mereghetti} S.,  2008, \aapr, 15, 225

\bibitem[\protect\citeauthoryear{{Mirabal} \& {Gotthelf}}{{Mirabal} \&
  {Gotthelf}}{2009}]{mirabal09}
{Mirabal} N.,  {Gotthelf} E.~V.,  2009, GCN Circ., 9696

\bibitem[\protect\citeauthoryear{{Morgan}, {Bloom} \& {Klein}}{{Morgan}
  et~al.}{2009}]{morgan09}
{Morgan} A.~N.,  {Bloom} J.~S.,    {Klein} C.~R.,  2009, GCN Circ., 9635

\bibitem[\protect\citeauthoryear{{Morris}, {Beardmore}, {Evans}, {Krimm},
  {Mangano}, {Mao}, {Markwardt}, {Page}, {Palmer}, {Rowlinson}, {Siegel},
  {Stark}, {Tagliaferri}, {Ukwatta} \& {Ziaeepour}}{{Morris}
  et~al.}{2009}]{morris09}
{Morris} D.~C. et~al.,  2009, GCN Circ., 9625

\bibitem[\protect\citeauthoryear{{Osborne}, {Beardmore}, {Evans} \&
  {Goad}}{{Osborne} et~al.}{2009}]{osborne09}
{Osborne} J.~P.,  {Beardmore} A.~P.,  {Evans} P.~A.,    {Goad} M.~R.,  2009,
  GCN Circ., 9636

\bibitem[\protect\citeauthoryear{{Reichart} \& {Price}}{{Reichart} \&
  {Price}}{2002}]{reichart02}
{Reichart} D.~E.,  {Price} P.~A.,  2002, \apj, 565, 174

\bibitem[\protect\citeauthoryear{{Rowlinson} \& {Morris}}{{Rowlinson} \&
  {Morris}}{2009}]{rowlinson09}
{Rowlinson} A.,  {Morris} D.~C.,  2009, GCN Circ., 9642

\bibitem[\protect\citeauthoryear{{Sakamoto}, {Barthelmy}, {Baumgartner},
  {Cummings}, {Fenimore}, {Gehrels}, {Krimm}, {Markwardt}, {Morris}, {Palmer},
  {Sato}, {Stamatikos}, {Tueller} \& {Ukwatta}}{{Sakamoto}
  et~al.}{2009}]{sakamoto09}
{Sakamoto} T. et~al.,  2009, GCN Circ., 9640

\bibitem[\protect\citeauthoryear{{Senziani}, {Novara}, {de Luca}, {Caraveo},
  {Belloni} \& {Bignami}}{{Senziani} et~al.}{2007}]{senziani07}
{Senziani} F.,  {Novara} G.,  {de Luca} A.,  {Caraveo} P.~A.,  {Belloni} T.,
  {Bignami} G.~F.,  2007, \aap, 476, 1297

\bibitem[\protect\citeauthoryear{Stratta et 
al.}{2004}]{stratta04} Stratta G., Fiore F., Antonelli L.~A., 
Piro L., De Pasquale M., 2004, ApJ, 608, 846 

\bibitem[\protect\citeauthoryear{{Vaughan}, {van der Klis}, {Wood}, {Norris},
  {Hertz}, {Michelson}, {van Paradijs}, {Lewin}, {Mitsuda} \&
  {Penninx}}{{Vaughan} et~al.}{1994}]{vaughan94}
{Vaughan} B.~A. et~al.,  1994, \apj, 435, 362

\bibitem[\protect\citeauthoryear{{von Kienlin}, {Beckmann}, {Rau}, {Arend},
  {Bennett}, {McBreen}, {Connell}, {Deluit} \& {et~al.}}{{von Kienlin}
  et~al.}{2003}]{vonkienlin03short}
{von Kienlin} A. et~al., 2003, \aap, 411, L299

\bibitem[\protect\citeauthoryear{{Wolfe}, {Gawiser} \& {Prochaska}}{{Wolfe}
  et~al.}{2005}]{wolfe05}
{Wolfe} A.~M.,  {Gawiser} E.,    {Prochaska} J.~X.,  2005, \araa, 43, 861

\end{thebibliography}
\bsp

\label{lastpage}

\end{document}